\documentclass[a4paper,fleqn,usenatbib]{mnras}

\usepackage{newtxtext,newtxmath}

\usepackage[T1]{fontenc}
\usepackage{ae,aecompl}

\usepackage{graphicx}	
\usepackage{amsmath}	

\usepackage{amssymb}	
\usepackage{color}

\title[RWI in PPDs with Cooling]{Rossby Wave Instabilities of 
			Protoplanetary Discs with Cooling}

\author[Huang \& Yu]{
Shunquan Huang$^{1,2,3}$
and Cong Yu$^{1,2,3}$\thanks{E-mail: yucong@mail.sysu.edu.cn}
\\
$^{1}$School of Physics and Astronomy, Sun Yat-Sen University, Zhuhai 519082, China\\
$^{2}$CSST Science Center for the Guangdong-Hong Kong-Macau Greater Bay Area, Zhuhai 519082, China\\
$^{3}$State Key Laboratory of Lunar and Planetary Sciences, Macau University of Science and Technology, Macau, China
}

\date{Accepted XXX. Received YYY; in original form ZZZ}

\pubyear{2021}

\begin{document}
\label{firstpage}
\pagerange{\pageref{firstpage}--\pageref{lastpage}}
\maketitle

\begin{abstract}
Rossby wave instabilities (RWIs) usually lead to nonaxisymmetric vortices in protoplanetary discs and some observed sub-structures of these discs can be well explained by RWIs. 
We explore how the cooling influences the growth rate of unstable RWI modes in terms of the linear perturbation analysis. 
The cooling associated with the energy equation is treated in two different ways. The first one we adopt is a simple cooling law. The perturbed thermal state relaxes to the initial thermal state on a prescribed cooling timescale. In the second, we treat the cooling as a thermal diffusion process.
The difference in the growth rate between the adiabatic and isothermal modes becomes more pronounced for discs with smaller sound speed.
For the simple cooling law, the growth rates of unstable modes monotonically decrease with the shorter cooling timescale in barotropic discs.
But the dependence of growth rate  with the cooling timescale becomes non-monotonic in non-baratopic discs. The RWI might even be enhanced in non-barotropic discs during the transition from the adiabatic state to the isothermal state.
When the cooling is treated as the thermal diffusion, even in barotropic disc, the variation of growth rate with thermal diffusivity becomes non-monotonic. Further more, a maximum growth rate may appear with an appropriate value of thermal diffusivity. 
The angular momentum flux (AMF) is investigated to understand the angular momentum transport by RWI with cooling. 
\end{abstract}




\section{Introduction}
Recently, ALMA observations have revealed the nonaxisymmetric dust distribution in 
protoplanetary discs
\citep{2013Sci...340.1199V, 2019ApJ...882...49L, 2020ApJ...892..111F, 2021AJ....161...33V}. 
These observations suggest that there might be large scale asymmetric substructures in discs, such as 
large scale vortices \citep{2014ApJ...783L..13P}. These large scale vortices may be formed by the merger of several small scale vortices \citep{2000ApJ...537..396G}.
Due to the gas drag, the dust particles are
captured in the vortices and accumulate to form planetesimals effectively
\citep{2010ApJS..190..297B, 1995A&A...295L...1B, 2000A&A...356.1089C, 2013ApJ...775...17L}. The trap of the dust and the formation of planetesimals caused by the vortices are of essential importance in the course of planet formation. 

A natural physical explanation for these vortices is the Rossby Wave Instability (RWI). The RWI was originally studied in two-dimensional (2D) accretion disc
\citep{1999ApJ...513..805L, 2000ApJ...533.1023L, 2010A&A...521A..25U, 
2016ApJ...823...84O}. These works established
that the discs with a steep radial structure, such as the surface density enhancement 
or gap, are unstable to 
nonaxisymmetric perturbations. 
\cite{2001ApJ...551..874L} showed the merger of RWI vortices by nonlinear 
hydrodynamic simulations, which is consistent with the results by \citet{2000ApJ...537..396G}
(also see \cite{2006ApJ...649..415I, 2006A&A...446L..13V, 2018ApJ...864...70O}). 
Later on, the RWI properties have been further examined under various circumstances, 
such as the dusty disc with particle accumulation 
\citep{2008A&A...491L..41L, 2009A&A...497..869L}, the disc with self-gravity
\citep{2018MNRAS.478..575L, 2018MNRAS.479.4878P},
and the disc with magnetic fileds \citep{2009ApJ...702...75Y}. Three-dimensional (3D) RWI calculations had been performed both with linear 
analyses \citep{2012MNRAS.422.2399M, 2012ApJ...754...21L} and non-linear numerical simulations
\citep{2010A&A...516A..31M}. 
3D calculations and 2D calculations showed similar results
\citep{2012ApJ...754...21L, 2013A&A...559A..30R}.
Therefore, it is physically reasonable for us to study the RWI properties using 2D assumptions for simplicity.

The RWI in protoplanetary discs with cooling has been widely studied
\citep{2015MNRAS.450.1503L,2015ApJ...810...94L,2020MNRAS.493.3014T,2021ApJ...922...13F}. All these studies are based on nonlinear numerical simulations. 
These works focus on the long term nonlinear evolution of the vortices excited by RWI. \cite{2015MNRAS.450.1503L} studied the RWI with cooling in the early linear stage and concluded that the mode growth rate decreases when the disc cools down more slowly. Numerical simulations have revealed that the variation of RWI growth rate with cooling are actually more complex \citep{2021ApJ...922...13F}. To further understand how the cooling affects the behavior of RWI, we consider the thermal response of RWI with two different ways of incorporating cooling into the energy equation. The first one is a simple cooling law. We assume that the perturbed thermal state relaxes to its initial equilibrium thermal state on a cooling timescale, $t_c$.
In the second way, we treat the cooling as a thermal diffusion process. Discs with thermal conduction have 
been widely discussed for the subcritical baroclinic instability 
\citep{2003ApJ...582..869K,2007ApJ...658.1236P,2007ApJ...658.1252P,2010A&A...513A..60L,2016A&A...592A.136B}, 
which is another robust mechanism to generate vortices. The thermally driven torques on protoplanet by the conduction have influential effects on the planet migration \citep{2008A&A...485..877P,2011MNRAS.410..293P,2017MNRAS.472.4204M,2020ApJ...902...50H}.
In this work, we perform global linear analysis of the RWI both with the simple cooling law and the cooling by thermal diffusion. 
We can get a comprehensive understanding of how the cooling influences the RWI by comparing the results of different treatment of cooling. 

The angular momentum transport in protoplanetary discs are of particular importance for planet formation \citep{1980ApJ...241..425G}. 
It is well established that the planet excites spiral density waves at the Lindblad resonances \citep{1978ApJ...222..850G,1995ARA&A..33..505P,2016ApJ...826...75D},
so that the angular momentum transports from the inner region to the outer region. The angular momentum exchange between the planet and the protoplanetary discs leads to the planet migration as well as the gap formation \citep{1986ApJ...309..846L,1996ApJ...460..832T,2002ApJ...565.1257T,2017ApJ...843..127D}. 
Prior studies had shown that the integrated angular momentum flux (AMF) of these density waves 
is conserved \citep{1979ApJ...233..857G}. 
But a recent study shows that some interesting behaviors of AMF appears for protoplanetary discs with cooling, especially for the isothermal protoplanetary disc 
\citep{2019ApJ...878L...9M,2020ApJ...892...65M}. 
Since the waves excited by RWI outside (inside) the outer (inner) 
Lindblad resonance are similar to those excited by planets, it would be interesting to investigate the AMF driven by RWI in discs with cooling.

This paper is structured as follows: we present the equilibrium disc setup  
in section \ref{sec:equilibrium_disc}. Basic equations and methods of solving the equations 
are shown in section \ref{sec:equations}. We display 
the numerical solutions to the perturbation equations in section \ref{sec:results}.  
Conclusions and discussions are generalized in section \ref{sec:conclution}.

\section{Equilibrium Disc}
\label{sec:equilibrium_disc}
We consider equilibrium discs in cylindrical $\left( r,\phi, z\right)$ coordinates, 
which are axisymmetric $\left( \partial/\partial\phi = 0\right)$ 
and in steady state $\left( \partial/\partial t = 0\right)$.
The vertical half-thickness is assumed to be much smaller than the radial
distance ($h\ll r$) so that the disc can be simplified as a two dimensional
disc. Both surface density $\Sigma(r)$ and pressure $P(r)$ are integrated 
vertically. The flow is rotating in a velocity 
$\boldsymbol{v}=v_\phi \hat{\phi}$, where $v_\phi$ is given by the force balance in the radial direction
\begin{equation}
	\frac{v_{\phi}^2}{r} \equiv r\Omega^2 
	= \frac{1}{\Sigma} \frac{dP}{dr} + \frac{d\Phi}{dr} \ .
	\label{force_balance}
\end{equation}
Here $\Omega$ is the angular velocity and $\Phi$ is the gravitational potential 
of the central objet. The disc self-gravity is neglected in this paper.

We consider an equilibrium surface density profile
with an Gaussian gap, which is
\begin{equation}
	\Sigma=\Sigma_b \cdot \left\{ 1-(\mathcal{H}-1)\exp{\left[-\frac{1}{2} 
		\left( \frac{r-r_0}{\Delta}\right)^2\right]} \right\} \ , 
	\label{density_profile}
\end{equation}
where $\Sigma_b = \Sigma_0\left(r/r_0\right)^{-p}$ is the background surface density with a power law index $p$. The subscript $b$ and $0$ represent the profiles of background disc and its
value at $r_0 = 1.0$, respectively. Specifically, we take $\mathcal{H} = 1.5$, 
$\Sigma_0=0.5$, $\Delta = 0.05r_0$, and $p = 1.0$.
We use an ideal equation of state (EoS)
\begin{equation}
	P=(\gamma-1)e \Sigma \ ,
	\label{Equation_of_state_e}
\end{equation}
where $\gamma$ is the adiabatic index, and $e$ is the specific internal energy.
For two dimensional discs, the equation of state can be written as
\begin{equation}
	P=\frac{c_s^2}{\gamma} \Sigma \ .
	\label{Equation_of_state_cs}
\end{equation}

\subsection{Equilibrium for Simple Cooling Law}
\begin{figure}
	\includegraphics[width=\columnwidth]{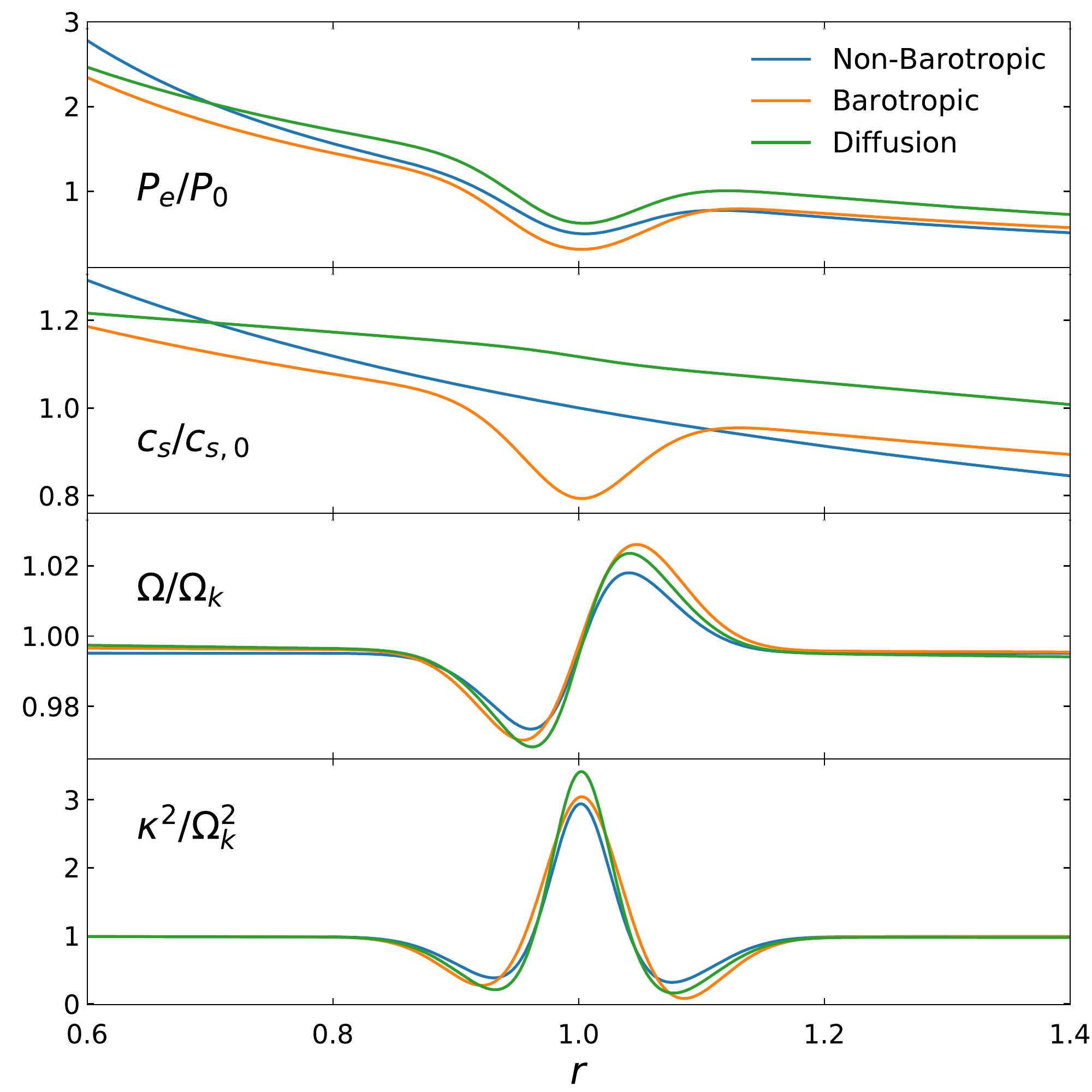}
	\caption{Equilibrium disc whose typical sound speed $c_{s,0}=0.09$.
			 From top to bottom, each panels display
			 pressure, $P/P_0$; sound speed, $c_s/c_{s,0}$; 
			 angular velocity, $\Omega/\Omega_k$; and square of epicyclic 
		 	 frequency $\kappa^2/\Omega_k^2$. Orange, blue and green lines
			 line represent barotropic, non-barotropic, and thermal equilibrium discs respectively.}
	\label{fig:equilibrium_disc}
\end{figure}

The sound speed profiles are specified for barotropic and non-barotropic discs as follows.
For barotropic discs, pressure are determined by density with the relationship of
${P}/{P_0} = \left({\Sigma}/{\Sigma_0}\right)^\gamma$,
where $P_0 = c_{s,0}^2\Sigma_0/\gamma$, and $c_{s,0}$ is a typical sound speed.
Typically, we take the sound speed, or equivalently the disc aspect ratio $H=c_{s,0}/v_{\phi,0}
=h/r $ in the range of $0.05\sim 0.1$.
The sound speed of the disc is determined according to $c_s=\sqrt{\gamma P/\Sigma}$.
For non-barotropic discs, we exploit an artificial sound speed profile
\begin{equation}
	c_s = c_{s,0}\left(r/r_0\right)^{-q/2},
	\label{nonbaro_cs}
\end{equation}
where $q=1.0$ is a constant. In Figure~\ref{fig:equilibrium_disc}, 
we show the equilibrium profiles of $P(r)/P_0$, $c_s(r)/c_{s,0}$,
$\Omega(r)/\Omega_k(r)$, and $\kappa^2(r)/\Omega_k^2(r)$ for barotropic and non-barotropic discs with a constant $c_{s,0} = 0.09$. Here
$\Omega_k = \left(\frac1r\frac{d\Phi}{dr}\right)^{1/2}$ is the rotation frequency of 
Keplerian discs, and $\kappa^2 \equiv \frac{1}{r^3} \frac{d(\Omega^2r^4)}{dr}$ 
is the square of radial epicyclic frequency.
In our calculation, the radii of the disc is in range of $0.4 \leq r \leq 1.6$.

\subsection{Equilibrium for Thermal Diffusion }
For discs with thermal diffusion, it is necessary to keep the thermal
equilibrium to make sure the disc is in steady state, i.e., 
\begin{equation}
	-\nabla\cdot\boldsymbol{F}_H = -\nabla\cdot\left(-\chi \Sigma\nabla e\right) = 0, 
	\label{diffusion_balance}	
\end{equation}
where $\boldsymbol{F}_H$ is the heat flux, and $\chi$ is the thermal diffusivity. 
Generally $\chi$ is a function of $r$, but for simplicity we assume it as a constant in this study.
Combining with the equation~(\ref{Equation_of_state_e}), the equation~(\ref{diffusion_balance}) becomes
\begin{equation}
	\frac{\partial^2 P}{\partial r^2} 
	+ \left(\frac1r - \frac{\Sigma^\prime}{\Sigma}\right)\frac{\partial P}{\partial r}
	+ \left(\frac{\left(\Sigma^\prime\right)^2}{\Sigma^2} 
			- \frac{\Sigma^{\prime\prime}}{\Sigma} - \frac{\Sigma^\prime}{r\Sigma}\right)P = 0, 
	\label{eq:diffusion_pressure}
\end{equation}
where the primes denote the radial derivatives. 
For a given $c_{s,0}$, we fix the boundary pressure using $P = c_{s,0}\left(r/r_0\right)^{-q/2} \left(\Sigma/\gamma\right)$ at both the inner and outer boundary of the disc. The equilibrium pressure $P$ can be solved from the equation~(\ref{eq:diffusion_pressure}) as a boundary value problem. 
Different from non-barotropic discs, constant $q$ is chosen to be $0.5$ and the radius range is $0.4 \leq r \leq 2.0$ here.
A representative profile for a thermal diffusion equilibrium disc is also shown as a green curve in Figure~\ref{fig:equilibrium_disc} with $c_{s,0} = 0.09$.

\subsection{Axisymmetric Stability}
\begin{figure}
	\includegraphics[width=\columnwidth]{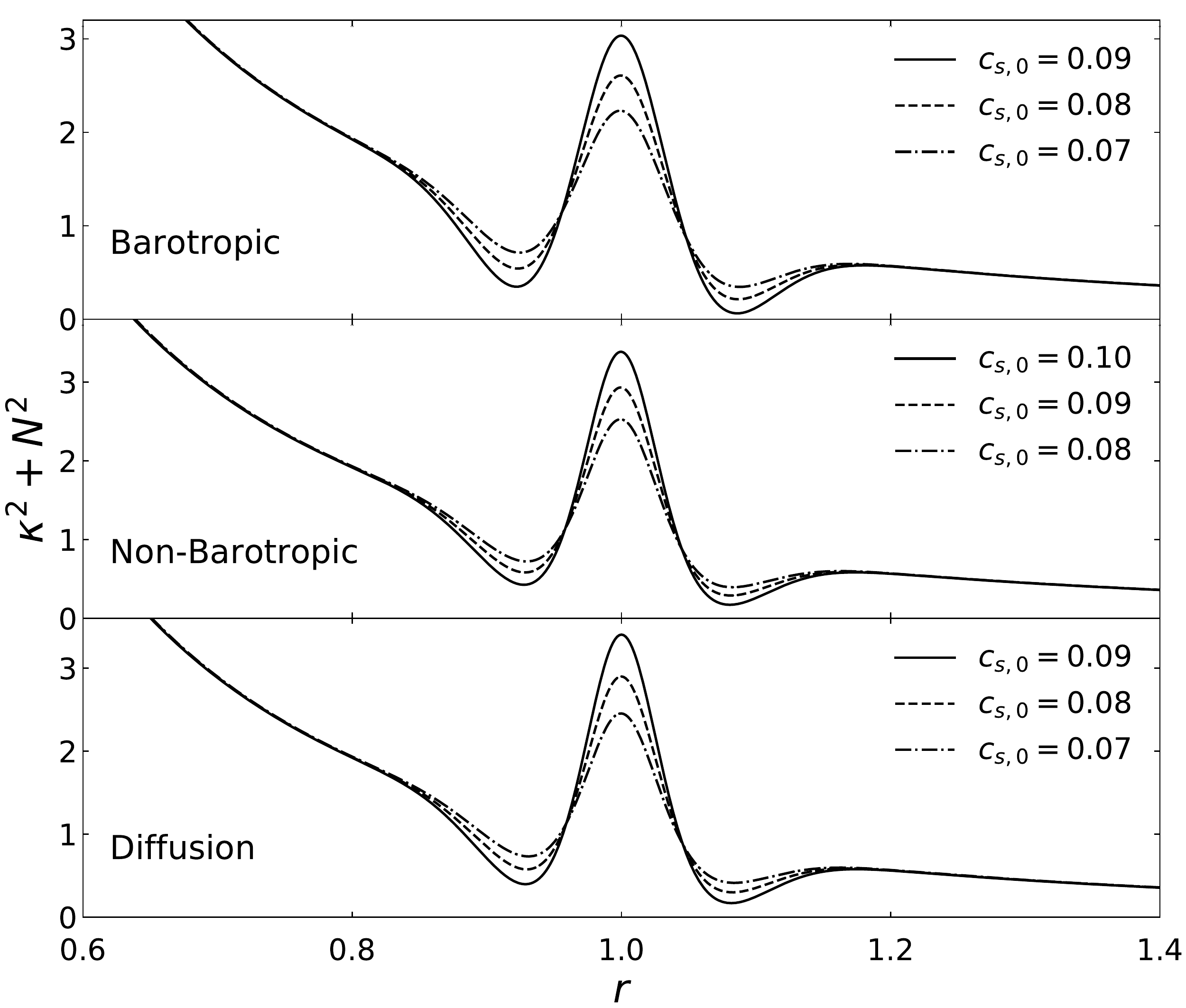}
	\caption{The profile of $\kappa^2+N^2$.
		The Upper, middle, and lower panel represent barotropic, non-barotropic, 
		and thermal equilibrium discs, respectively.
		Solid line, dashed line and dot-dashed line
		represent $c_{s,0}=0.09$, $0.08$, and $0.07$ for barotropic and thermal 
		equilibrium discs, and $c_{s,0}=0.10$, $0.09$, and $0.08$ for non-barotropic 
		discs, respectively. Note that these profiles are all positive.}
	\label{fig:Axisymmetric_stability}
\end{figure}

To investigate the non-axisymmetric instability of the discs, it is necessary to ensure that 
the equilibrium states we construct are stable to axisymmetric perturbations. 
The well known sufficient condition for local stability is the Solberg-Høiland 
criterion \citep{1936...Solberg.H,1939...Hoiland.E}, which reads \citep{1978ApJ...220..279E}
\begin{equation}
	\kappa^2(r)+N^2(r) \ge 0 \ ,
	\label{Axi_stab_crit}
\end{equation}
where 
\begin{equation}
	N^2 \equiv \frac{1}{\Sigma}\frac{dP}{dr}\left(\frac{1}{\Sigma}\frac{d\Sigma}{dr}
	-\frac{1}{\gamma P}\frac{dP}{dr}\right)
\end{equation}
is the radial Brunt-V{\"a}is{\"a}l{\"a} frequency. For barotropic discs,
we can easily see that $N^2\rightarrow 0$. For non-barotropic discs, 
$N^2$ is not negligible and $N^2$ remains much smaller than $\kappa^2$.
The profiles of $\kappa^2+N^2$ with different values of $c_{s,0}$
are displayed in Figure \ref{fig:Axisymmetric_stability}. 

\section{Linear Analysis with Different Energy Equations}
\label{sec:equations}
We now consider the linear perturbations to the equilibrium discs. The surface density, gas pressure and the flow velocity can be viewed as the sum of the equilibrium state and the perturbation state,
$\widetilde\Sigma=\Sigma+\delta\Sigma$,
$\widetilde P = P + \delta P$ and $\boldsymbol{\widetilde v} 
= \boldsymbol{v}+\delta \boldsymbol{v}$, respectively, where the 
prefix $\delta$ represent the perturbed values and $\delta \boldsymbol{v} 
= \left( \delta v_r, \delta v_\phi, 0 \right)$. The two-dimensional
mass and momentum conservation equations for discs are
\begin{equation}
	\frac{D\widetilde{\Sigma}}{Dt}+\widetilde{\Sigma} \nabla \cdot
	\boldsymbol{\widetilde v} = 0 \ ,
	\label{equation_of_Euler_mass}
\end{equation}
\begin{equation}
	\frac{D\boldsymbol{\widetilde v}}{Dt} 
	= -\frac{1}{\widetilde{\Sigma}} \nabla \widetilde{P}
	-\nabla \Phi \ ,
	\label{equation_of_Euler_moventum}
\end{equation}
where $D/Dt \equiv \partial/\partial t + \boldsymbol{\widetilde v} \cdot \nabla$.
The perturbation variables can be expressed in the form of
$ f(r) \exp(im\phi - i\omega t) $, where $m$ is an integer and 
$\omega = \omega_r + i\omega_i$ is the mode eigen-frequency. The continuity equations can be linearized as follows,
\begin{equation}
	i\sigma \delta \Sigma = \left(\Sigma^\prime + \frac{\Sigma}{r} \right)\delta v_r
	+ i k_\phi \Sigma \delta v_\phi + \Sigma\frac{\partial \delta v_r}{\partial r},
	\label{equation_of_perturbation_density}
\end{equation}
where $\sigma = \omega - m\Omega$ is the Doppler-shifted frequency, 
and $k_\phi=m/r$ is the azimuthal wavenumber. 
The linearized momentum equations can be written as
\begin{equation}
	i\sigma\delta v_r + 2\Omega\delta v_\phi = \frac{1}{\Sigma}
	\frac{\partial\delta P}{\partial r} - \frac{\delta\Sigma}{\Sigma^2}P^\prime \ ,
	\label{equation_of_perturbation_v_r}
\end{equation}
\begin{equation}
	i\sigma\delta v_\phi - \frac{\kappa^2}{2\Omega}\delta v_r = 
	ik_\phi\frac{\delta P}{\Sigma} \ .
	\label{equation_of_perturbation_v_phi}
\end{equation}

\subsection{Perturbation Equations for Discs with Simple Cooling}
\label{sec:per_equa_cooling}
\cite{2007prpl.conf..607D} indicates that there are two methods to treat the disc cooling. One method is called the simple cooling laws, which reads,
\begin{equation}
	\frac{D\widetilde{e}}{Dt} + \widetilde{P} \frac{D}{Dt} 
	\left(\frac{1}{\widetilde{\Sigma}}\right)
		= -\frac{\widetilde{e}-e}{t_c} \ .
	\label{equation_of_energy}
\end{equation}
\citet{2020ApJ...892...65M} also adopted such a method to include cooling in the energy equations. 
If we set the 
right hand side of equation~(\ref{equation_of_energy}) to be $0$, 
it reduces to $D(\widetilde{P}/\widetilde{\Sigma}^\gamma)/Dt = 0$, which becomes the adiabatic equation \citep{1995ARA&A..33..505P}.
The right hand side of equation (\ref{equation_of_energy}) is the cooling term. 
Note that $\widetilde{e}=\left.\widetilde{P} \middle/ \left[(\gamma -1)
\widetilde{\Sigma}\right]\right.$ (equilibrium $+$ perturbation) and
$e=\left[ P \middle/ \left[(\gamma -1)\Sigma\right] \right.$ (equilibrium). 
In the cooling term, the parameter $t_c$ is the cooling timescale on which 
the disc evolves towards the thermal equilibrium state. Different from
several previous studies setting $t_c\Omega=constant$,
we take $t_c=constant$ everywhere and define $\beta = t_c\Omega(r_0)$
as the dimensionless cooling parameter. 
Obviously, the discs become adiabatic or isothermal when 
$\beta\rightarrow\infty$ or $\beta\rightarrow 0$, respectively \citep{2020ApJ...892...65M}.

By linearizing the equation~(\ref{equation_of_energy}), we have 
\begin{equation}
	\left( \frac{1}{t_c}-i\sigma \right)\delta P 
	- \left( \frac{1}{\gamma t_c}-i\sigma \right)c_s^2\delta\Sigma
	+ \frac{\Sigma c_s^2}{L_S}\delta v_r = 0 \ .
	\label{equation_of_perturbation_cooling}
\end{equation}
Here we define a length scale of entropy variation $L_S$ as
\begin{equation*}
	\frac{1}{L_S} = \frac{1}{\gamma}\frac{d}{dr} 
	\left[\ln\left( \frac{P}{\Sigma^\gamma}\right) \right].
	\label{L_S}
\end{equation*}
To make the following linearized equations in a more compact form, we also define the length scale of pressure variation $L_P$ and the length scale of surface density variation $L_\Sigma$ as
\begin{equation*}
	\frac{1}{L_P} = \frac{1}{\gamma P}\frac{dP}{dr} \ , \quad
	\frac{1}{L_\Sigma} = \frac{1}{\Sigma}\frac{d\Sigma}{dr} \ ,
	\label{L_P_L_Sigma}
\end{equation*}
so that $\left(L_S\right)^{-1}=\left(L_P\right)^{-1}-\left(L_\Sigma\right)^{-1}$.
With some tedious mathematical manipulations, 
equations~(\ref{equation_of_perturbation_density})-(\ref{equation_of_perturbation_v_phi})
and (\ref{equation_of_perturbation_cooling}) can be cast into two equations for 
$\Psi = \delta P/\Sigma$ and $\delta v_r$, viz.,  
\begin{equation}
	\frac{\partial \Psi}{\partial r} + A_{11} \Psi + A_{12} \delta v_r = 0 \ ,
	\label{equation_Master_Psi}
\end{equation}
\begin{equation}
	\frac{\partial \delta v_r}{\partial r} + A_{21} \Psi + A_{22} \delta v_r = 0 \ ,
	\label{equation_Master_delta_v_r}
\end{equation}
where the four coefficients are
\begin{equation}
	A_{11} = \frac{1}{L_\Sigma} - \frac{2\Omega}{\sigma}k_\phi  
			- \frac{1}{L_P}\left(\frac{\gamma-i\sigma\gamma t_c}
			{1-i\sigma\gamma t_c}\right),
	\label{A_11}
\end{equation} 
\begin{equation}
	A_{12} = -i\left[
			\sigma - \frac{\kappa^2}{\sigma} 
			- \frac{ic_s^2}{L_PL_S}\left(\frac{\gamma t_c}
			{1-i\sigma\gamma t_c}\right)
			\right],
	\label{A_12}
\end{equation} 
\begin{equation}
	A_{21} = i\left[
			\frac{k_\phi^2}{\sigma}
			- \frac{\sigma}{c_s^2}\left(\frac{\gamma-i\sigma\gamma t_c}
			{1-i\sigma\gamma t_c}\right)
			\right],
	\label{A_21}
\end{equation} 
and 
\begin{equation}
	A_{22} = \left[
			\frac{1}{r} + \frac{1}{L_\Sigma} 
			+ \frac{\kappa^2 k_\phi}{2\Omega\sigma}
			-\frac{i\sigma}{L_S}\left(\frac{\gamma t_c}
			{1-i\sigma\gamma t_c}\right)
			\right].
\end{equation}
Note that the equations~(\ref{equation_Master_Psi}) and (\ref{equation_Master_delta_v_r})
determine the eigenfrequency $\omega$ and the relevant wave functions. 
Note that the imaginary part of the eigenfrequency, $\omega_i$, is the growth rate of unstable modes \footnote{$\omega_i\le 0$ for the stable modes and $\omega_i>0$ for the unstable modes.}.
In the above coefficients, there are four terms 
involving $t_c$. Two of them depend on $L_S$. These two terms will vanish if 
the disc is barotropic ($|L_S|\rightarrow\infty$).
Thus, we would expect that the numerical result will be a little different between 
the barotropic and non-barotropic discs. 
Also note that equations~(\ref{equation_Master_Psi}) and (\ref{equation_Master_delta_v_r})
can be combined as a single second order differential equation of $\Psi$,
\begin{equation}
	\Psi^{\prime\prime} + A_{31}(r)\Psi^\prime + A_{32}(r)\Psi = 0,
	\label{equation_Master_Loveplace}
\end{equation}
where
\begin{equation}
	A_{31}(r) = A_{11} + A_{22} - \frac{A_{12}^\prime}{A_{12}},
	\label{B(r)}
\end{equation}
\begin{equation}
	A_{32}(r) = A_{11}^\prime + A_{11}A_{22} - A_{12}A_{21} 
			- \frac{A_{11}A_{12}^\prime}{A_{12}}.
	\label{C(r)}
\end{equation}
When taking $\beta \rightarrow \infty$ for adiabatic discs, 
equation~(\ref{equation_Master_Loveplace}) reduces to equation~(10) 
in \cite{1999ApJ...513..805L}.

\subsection{Perturbation Equations for Discs with Thermal Diffusion}
When the cooling is treated as a thermal diffusion process, the energy equation can be written as 
\begin{equation}
	\frac{D\widetilde{e}}{Dt} + \widetilde{P} \frac{D}{Dt} \left(\frac{1}{\widetilde{\Sigma}}\right)
		= - \frac{1}{\widetilde{\Sigma}}\nabla\cdot\widetilde{\boldsymbol{F}_H} \ , 
	\label{equation_of_diffusivity}
\end{equation}
where $\widetilde{\boldsymbol{F}_H} = -\chi \widetilde{\Sigma}\nabla\widetilde{e}$.
Compared to equation~(\ref{equation_of_energy}), one can easily identify that $\chi$
is inversely proportional to $\beta$ (or $t_c$). The discs become 
adiabatic or isothermal when $\chi\rightarrow 0$ or $\chi\rightarrow\infty$, respectively.
Linearizing this equation with a constant $\chi$, we obtain the following perturbed energy equation, 
\begin{equation}
	\begin{aligned}
	\frac{P}{\Sigma}\frac{\partial^2\delta\Sigma}{\partial r^2} - \frac{\partial^2\delta P}{\partial r^2}
	& + B_1\frac{\partial\delta P}{\partial r} + B_2 \delta P   \\
	& + B_3 \frac{\partial\delta\Sigma}{\partial r} + B_4\delta\Sigma 
	  + \frac{c_s^2\Sigma}{\chi L_S}\delta v_r = 0 \ , 
	\label{equation_per_diff}
	\end{aligned}
\end{equation}
where 
\begin{equation}
	B_1 = \frac{1}{L_\Sigma} - \frac1r \ ,
	\label{B_1}
\end{equation}
\begin{equation}
	B_2 = k_\phi^2 - \frac{1}{L_\Sigma}\left(\frac{1}{L_\Sigma} - \frac1r\right)
		  +\frac{\Sigma^{\prime\prime}}{\Sigma} - i\frac{\sigma}{\chi} \ ,
	\label{B_2}
\end{equation}
\begin{equation}
	B_3 = c_s^2 \left( \frac{1}{\gamma r} + \frac{1}{L_P} - \frac{2}{\gamma L_\Sigma} \right) \ ,
	\label{B_3}
\end{equation}
and 
\begin{equation}
	B_4 = \frac{c_S^2}{\gamma L_\Sigma}\left( \frac{2}{L_\Sigma} - \frac1r 
			- \frac{\gamma}{L_P} - k_\phi^2L_\Sigma 
			- \frac{\Sigma^{\prime\prime}L_\Sigma}{\Sigma} \right)
		  + i\frac{\sigma c_s^2}{\chi} \ .
	\label{B_4}
\end{equation}
Note that,  by eliminating $\delta v_r$ and $\delta v_\phi$,
equations~(\ref{equation_of_perturbation_density})-(\ref{equation_of_perturbation_v_phi})
can be rearranged as a single equation for $\delta P$ and $\delta \Sigma$, which reads   
\begin{equation}
	\frac{\partial^2\delta P}{\partial r^2} 
		+ C_{11}\frac{\partial\delta P}{\partial r} + C_{12}\delta P 
		+ C_{13}\frac{\partial\delta \Sigma}{\partial r} + C_{14}\delta \Sigma = 0 \ , 
	\label{equation_Master_P}
\end{equation}
where the coefficients are 
\begin{equation}
	C_{11} = \frac{\sigma^\prime}{\sigma} - \frac{D^\prime}{D} + \frac1r 
			 - \frac{k_\phi\left(4\Omega^2 - \kappa^2\right)}{2\Omega\sigma}, 
	\label{C_11}
\end{equation}
\begin{equation}
	C_{12} = -k_\phi^2 - \frac{2\Omega k_\phi}{\sigma}
					\left(\frac{\Omega^\prime}{\Omega} - \frac{D^\prime}{D}\right),
	\label{C_12}
\end{equation}
\begin{equation}
	C_{13} = -\frac{c_s^2}{L_P},
	\label{C_13}
\end{equation}
\begin{equation}
	C_{14} = -D - \frac{c_s^2}{L_P}\left( \frac{\sigma^\prime}{\sigma} - \frac{D^\prime}{D}
					+ \frac{P^{\prime\prime}}{p^\prime} - \frac{1}{L_\Sigma} + \frac1r 
					+\frac{k_\phi \kappa^2}{2\Omega\sigma} \right),  
	\label{C_14}
\end{equation}
and 
\begin{equation*}
	D = \kappa^2 - \sigma^2.
\end{equation*}
By using equations~(\ref{equation_of_perturbation_v_r}) and (\ref{equation_of_perturbation_v_phi}) to eliminate $\delta v_r$, and equation~(\ref{equation_Master_P}) to eliminate 
${\partial^2\delta P}/{\partial r^2}$, we obtain from equation~(\ref{equation_per_diff})
\begin{equation}
	\frac{\partial^2\delta\Sigma}{\partial r^2} 
		+ C_{21}\frac{\partial\delta P}{\partial r} + C_{22}\delta P 
		+ C_{23}\frac{\partial\delta \Sigma}{\partial r} + C_{24}\delta \Sigma = 0, 
	\label{equation_Master_Sigma}
\end{equation}
where 
\begin{equation}
	C_{21} = \frac{\gamma}{c_s^2}\left(C_{11} + B_{1}\right) + i\frac{\gamma\sigma}{\chi DL_S},
	\label{C_21}
\end{equation}
\begin{equation}
	C_{22} = \frac{\gamma}{c_s^2}\left(C_{12} + B_{2}\right) - i\frac{2\gamma k_\phi\Omega}{\chi DL_S},
	\label{C_22}
\end{equation}
\begin{equation}
	C_{23} = \frac1r - \frac{2}{L_\Sigma},
	\label{C_23}
\end{equation}
\begin{equation}
	C_{24} = \frac{\gamma}{c_s^2}\left(C_{14} + B_{4}\right) 
			 - i\frac{\gamma\sigma c_s^2}{\chi DL_SL_P}.
	\label{C_24}
\end{equation}

\subsection{Methods of Solving Linear Eigenvalue Equations}
When we adopt the simple cooling law, the governing equations for the linear perturbations are Equations~(\ref{equation_Master_Psi}) and (\ref{equation_Master_delta_v_r}). They
consist of a two-point boundary eigenvalue problems, which can be solved readily by 
the relaxation method \citep{1992nrfa.book.....P}. In this method, we use approximate finite-difference 
equations (FDEs) on a mesh of points to replace the ordinary difference equations (ODEs). 
Then with an initial trial and through iteration, we obtain eigenvalue and eigenfunctions 
progressively by a method analogous to Newton-Raphson method. The computation cost of this method is 
very low so we can afford relatively high resolution and accuracy. Typically, we use 1001 uniform 
mesh points and the average error of each point is lower than $10^{-8}$.
When we treat the cooling as a thermal diffusion process, we numerically solve Equations~(\ref{equation_Master_P}) and (\ref{equation_Master_Sigma}) instead. Although these two equations are more complex, involving higher order derivatives than equations~(\ref{equation_Master_Psi}) and (\ref{equation_Master_delta_v_r}), they can be solved essentially in the same way\footnote{This relaxation method is also adopted to solve equation~(\ref{eq:diffusion_pressure})
to get the pressure distribution of thermal diffusion equilibrium.}.
And the boundary conditions we used is a WKB relation that require the group velocity of density waves 
propagate away from the central region in both the inner and outer parts of 
the disc \citep{2000ApJ...533.1023L}.

\section{Results} 
\label{sec:results}
In this section, we first show the general properties of RWI with cooling for the simple cooling law, and then we focus on how the cooling influences the variations of growth rate of unstable modes. Both the simple cooling law and the cooling as the thermal diffusion are discussed.
We finally present the behavior of the angular momentum flux (AMF) driven by the RWI modes.

\subsection{General Solutions with $m=5$ and $\beta =10^4$}
\label{sec:General_solu}
\begin{figure}
	\includegraphics[width=\columnwidth]{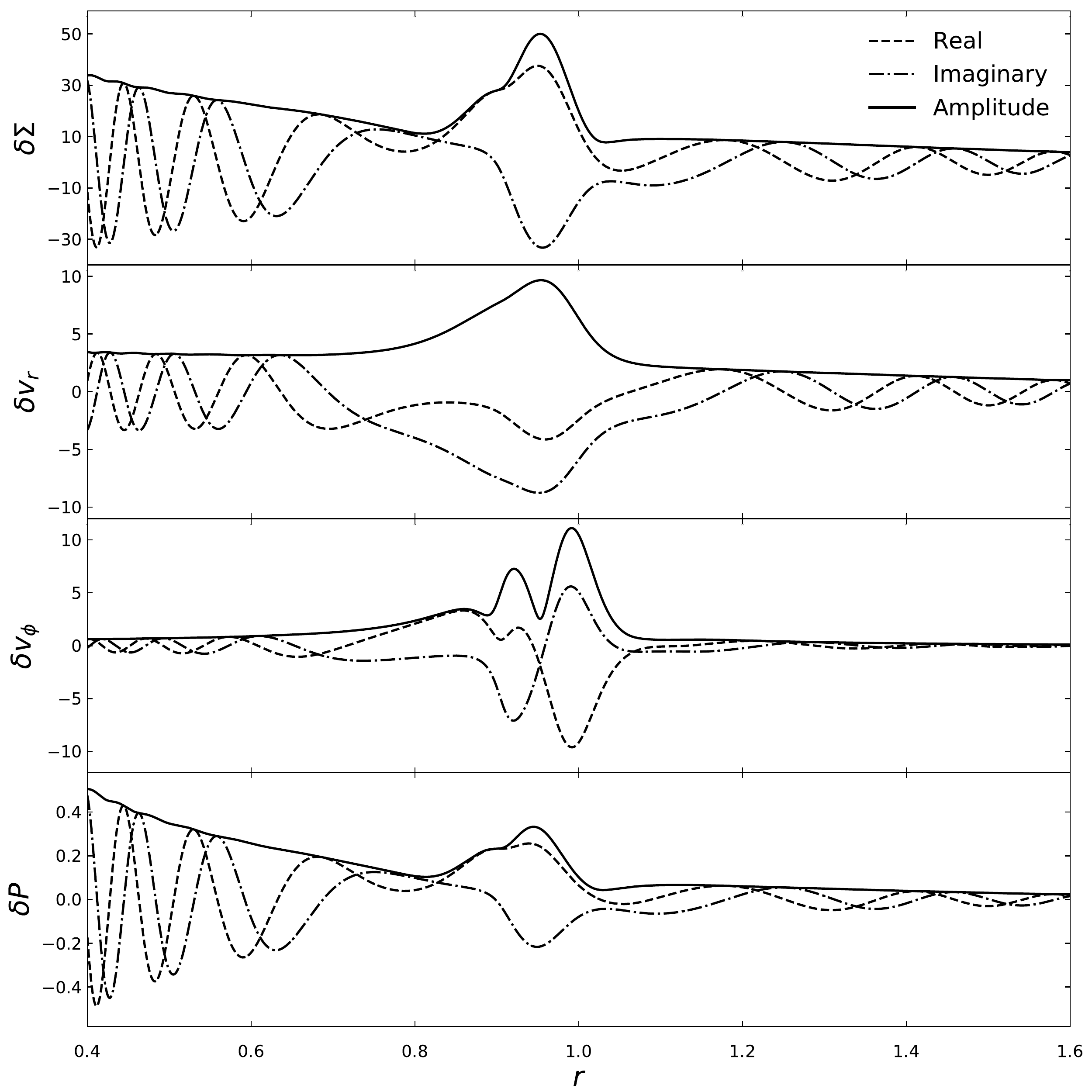}
	\caption{RWI eigenfunctions for the inner edge mode, displaying the perturbed
		density, the perturbed radial and azimuthal velocity, and the pressure 
		perturbation for $m=5$ and $\beta=10^4$. The dashed line, dot-dashed 
		line and solid line are the real part, the imaginary part and the amplitude, 
		respectively.}
	\label{fig:inner_eigenfunction}
\end{figure}

\begin{figure}
	\includegraphics[width=\columnwidth]{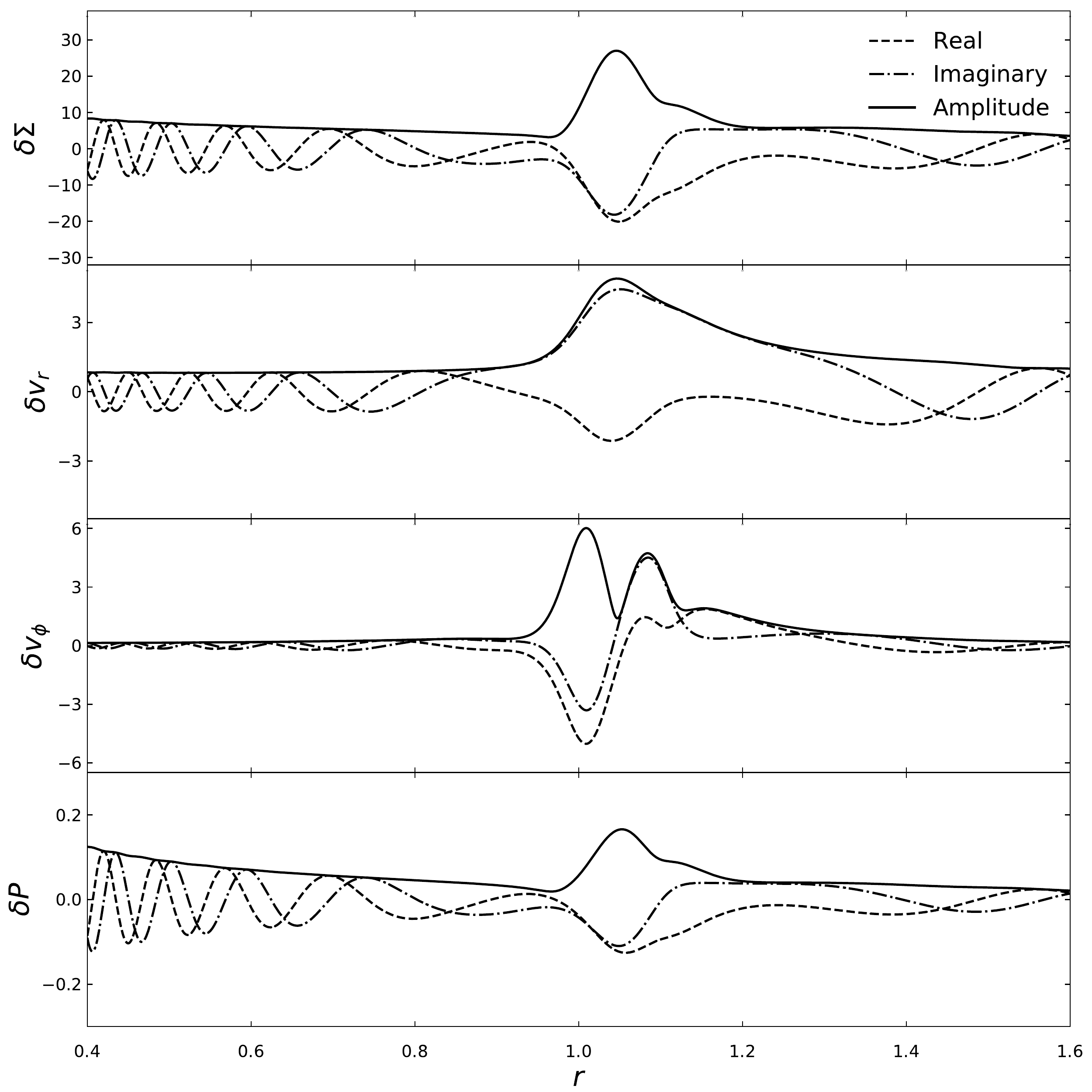}
	\caption{Same as Figure \ref{fig:inner_eigenfunction} but for the outer edge mode.}
	\label{fig:outer_eigenfunction}
\end{figure}

\begin{figure}
	\includegraphics[width=\columnwidth]{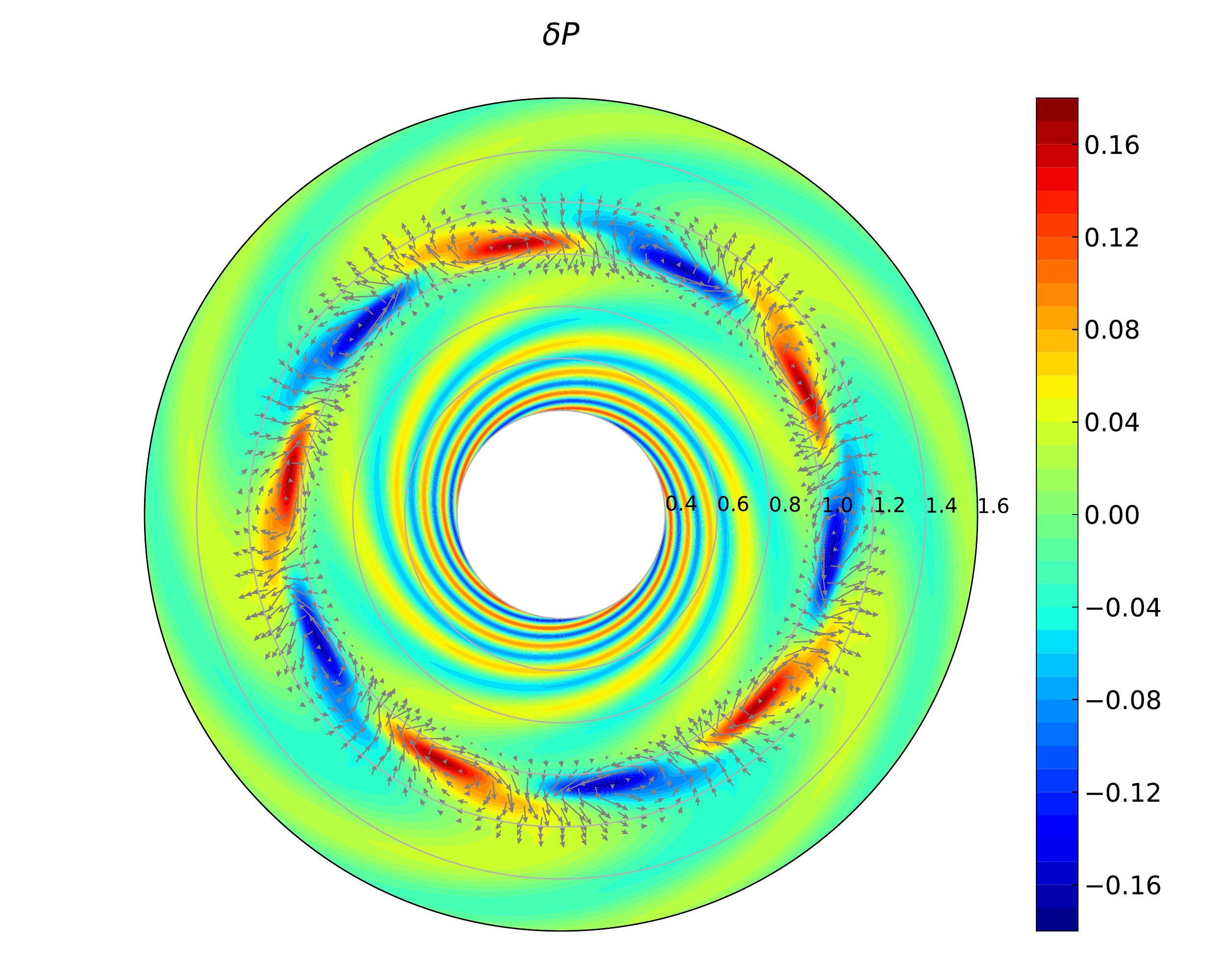}
	\caption{Two dimensional distribution of perturbed pressure in 
			 barotropic discs of outer edge modes with $m=5$ and $\beta = 10^4$.
			 Perturbed velocity is indicated by arrows.}
	\label{fig:2D_perturbations}
\end{figure}

For the discs surface density distribution with a Gaussian gap profile, we find two independent unstable
modes since there are two edges (the inner and outer edge) associated with the Gaussian gap. 
When the gap is deep enough, the potential vorticity (PV) associated with the two edges make the disc unstable \citep{2000ApJ...533.1023L}. 
The unstable eigenfunctions with the simple cooling law for barotropic discs 
are shown in Figure~\ref{fig:inner_eigenfunction} and~\ref{fig:outer_eigenfunction} 
for the inner and outer edge mode, respectively. The azimuthal wave number of these modes $m=5$ and the cooling parameter $\beta =10^4$. 
The inner edge mode has a growth rate $\omega_i/\Omega\left(r_0\right)=0.1776$
and a real frequency $\omega_r/\left[m\Omega\left(r_0\right)\right]=1.1479$,
while the outer edge mode has a growth rate $\omega_i/\Omega\left(r_0\right)=0.1845$
and a real frequency $\omega_r/\left[m\Omega\left(r_0\right)\right]=0.8671$.
\cite{2008MNRAS.387..446T} and \cite{2009MNRAS.393..979L} have shown
that Rossby wave zone lies inside the co-rotation radius where $\omega_r-\Omega=0$ 
and the density waves are launched at the Lindblad resonances where 
$\left(\omega_r-\Omega\right)=\pm\kappa$. In this case for outer edge mode, 
we find a co-rotation radius at $r\approx 1.11$, the inner Lindblad resonances at 
$r\approx 0.95$ and the outer Lindblad resonances at $r\approx 1.24$. We display the two dimensional distribution of outer edge mode in Figure~\ref{fig:2D_perturbations}, which shows the pressure perturbation distribution. It is clear that 
inner and outer Lindblad resonances represent the demarcations of vortices excited by RWI and density waves propagating away. Inner edge mode shows similar features.


\subsection{Unstable Growth Rates with Simple Cooling Law}
\begin{figure}
	\includegraphics[width=\columnwidth]{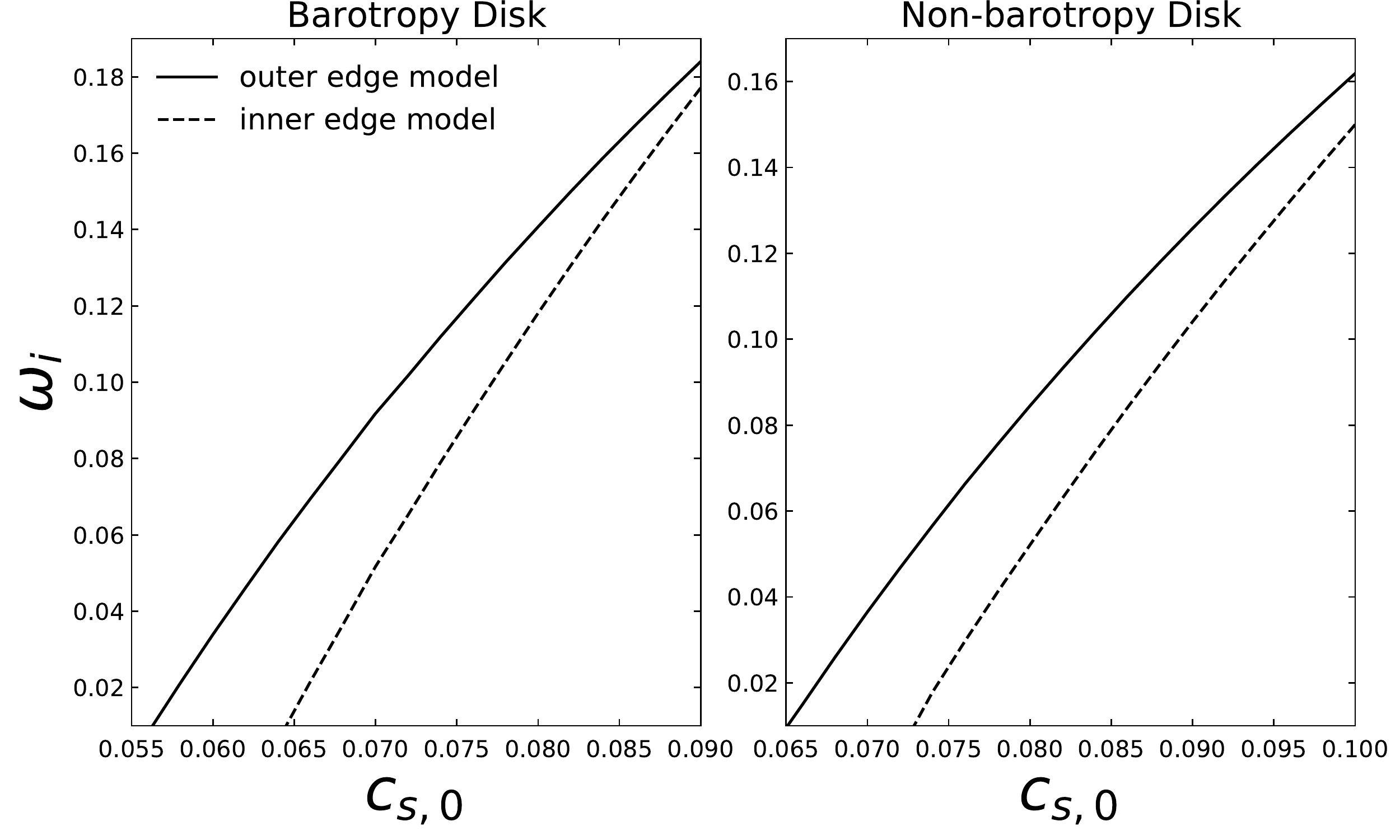}
	\caption{RWI growth rate variation with different $c_{s,0}$ for $\beta=10^4$,
			 where left side is for barotropic discs and right side is for 
			 non-barotropic discs. Also solid line represents the outer edge mode 
			 and dashed line represents the nner edge mode.}
	\label{fig:omega_i_cs}
\end{figure}

Here we show how the sound speed $c_{s,0}$ and the cooling parameter $\beta$ influence the variations of the growth rate, $\omega_i$, for the discs with the simple cooling law.
In Figure~\ref{fig:omega_i_cs}, we show the variation of $\omega_i$ with the sound speed $c_{s,0}$.
The mode growth rates decrease with the smaller $c_{s,0}$. This means that there exists a lower threshold of $c_{s,0}$, beneath which the RWI modes will be completely suppressed.
As mentioned above, there also exists the upper threshold of $c_{s,0}$ to keep the disc stable against the axisymmetric perturbations.
It is interesting to note that the growth rate of outer edge mode remains larger than the inner edge mode, indicating that outer edge mode is generally the dominant mode. 

\begin{figure}
	\includegraphics[width=\columnwidth]{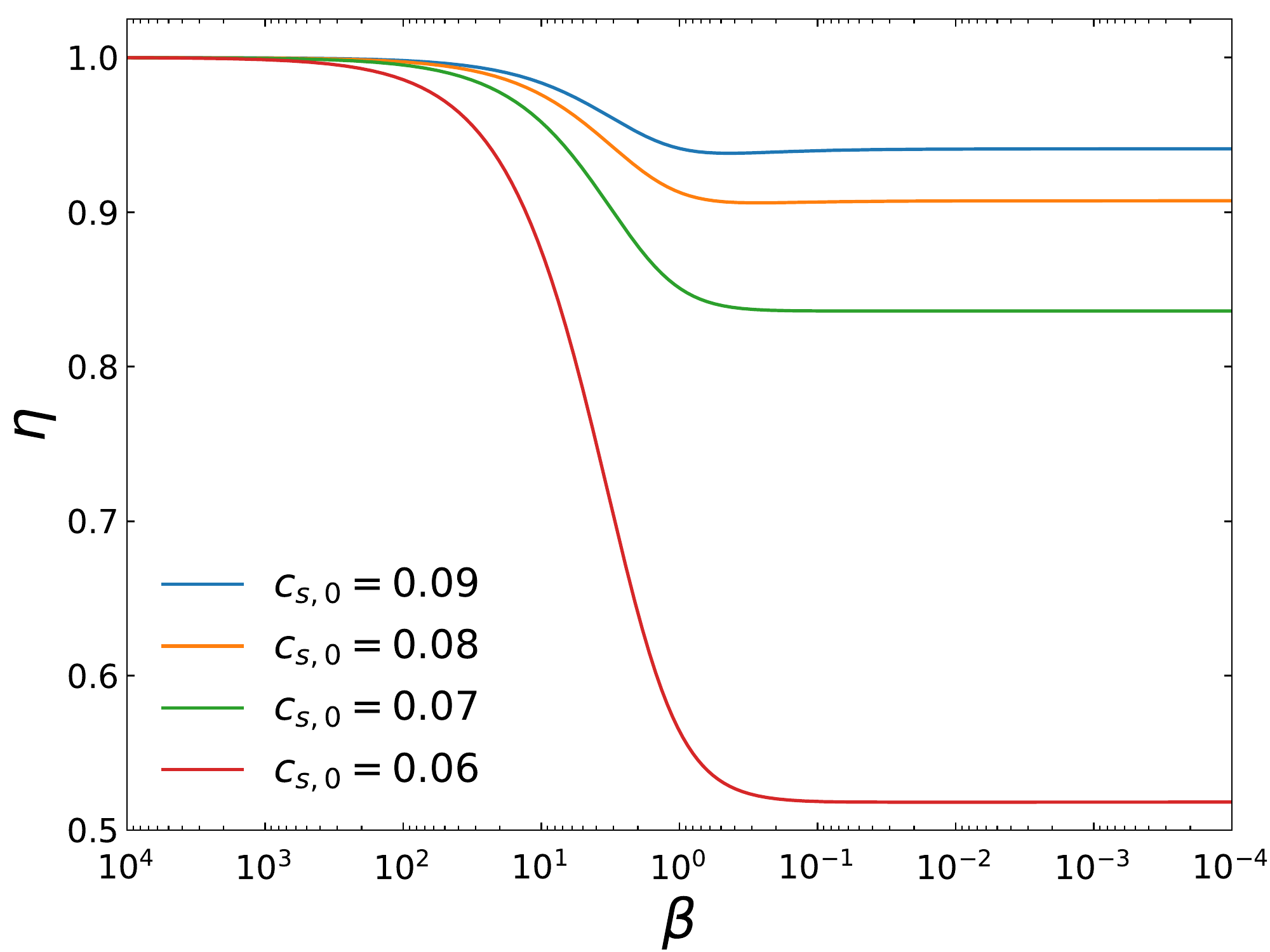}
	\caption{The profile of normalized outer edge mode growth rate $\eta$ againsts 
		dimensionless cooling timescale $\beta$ for barotropic discs. Here, the blue, 
		orange, green and red lines represent $c_{s0}=0.09$, $0.08$, $0.07$, 
		and $0.06$, respectively.}
	\label{fig:barotropic_cooling}
\end{figure}

In Figure~\ref{fig:barotropic_cooling}, we present the variations of the growth rate of outer edge modes with the cooling parameter $\beta$ for barotropic discs. Four different cases of sound speed $c_{s,0}$ are shown with different colors. 
We have normalized $\omega_i$ in terms of its value at $\beta = 10^4$ and introduce\footnote{Note that when $\beta$ is sufficiently large, the energy equation reduces to the adiabatic equation.} 
\begin{equation}
	\eta = \frac{\omega_i}{\omega_i\big|_{\beta=10^4}}.
	\label{normalization_beta}
\end{equation}
In Figure~\ref{fig:barotropic_cooling}, we find that the variations of $\eta$ with the cooling parameter $\beta$
is monotonic for barotropic discs, higher growth rate for adiabatic discs and lower for isothermal discs. The transition from the adiabatic disc to the isothermal disc occurs around $
\beta \sim 1$. It is clear that for barotropic discs, the RWI is always suppressed by the cooling. 
The suppression of the RWI $\eta$ appears to be more remarkable for the discs with the lower sound speed. 

\begin{figure}
	\includegraphics[width=\columnwidth]{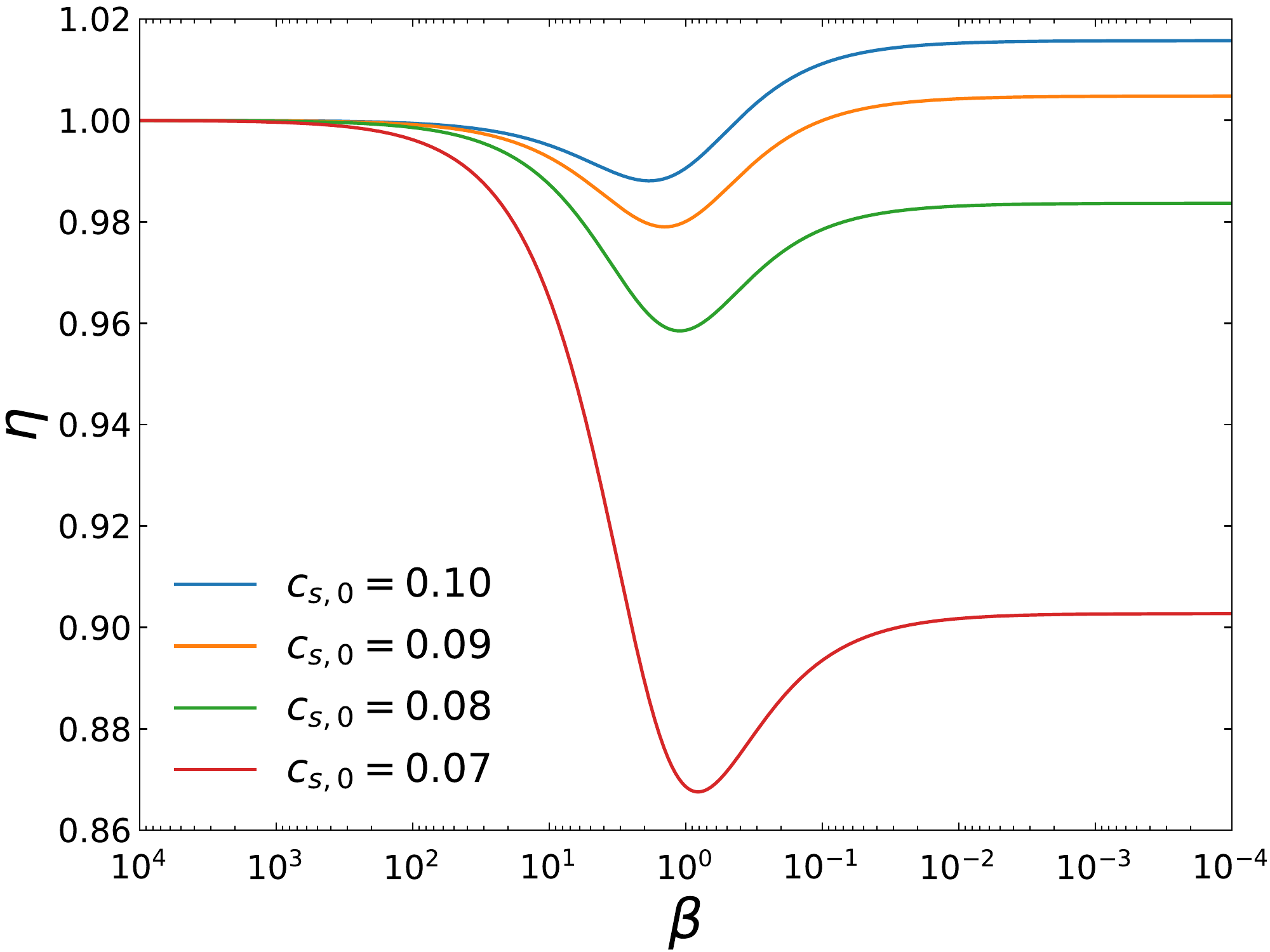}
	\caption{Same as Figure~\ref{fig:barotropic_cooling} but for non-barotropic discs. 
		Here, the blue, orange, green and red lines represent 
		$c_{s0}= 0.10$, $0.09$, $0.08$, and $0.07$, respectively.}
	\label{fig:non-barotropic_cooling}
\end{figure}

However, things become more complex when the discs are non-barotropic.  The behavior of 
$\eta$ with the cooling parameter $\beta$ is no longer monotonic. 
Figure~\ref{fig:non-barotropic_cooling} shows the curves of $\eta$ for non-barotropic discs. In
Figure~\ref{fig:non-barotropic_cooling},
we find a minimum of the growth rate at around $\beta \sim 1$. It is clear that this minimum increases with the sound speed, $c_{s,0}$. If the cooling parameter $\beta$ becomes smaller than $\sim 1$ (i.e., more efficient cooling),  the 
growth rate increases, and saturates at the isothermal limit. In sharp contrast to barotropic discs, for non-barotropic discs with high sound speed $c_{s,0}$,
the value of $\eta$ for sufficiently small value of $\beta$ (i.e., the isothermal limit) may even be enhanced compared to the value of $\eta$ at adiabatic limit (i.e., sufficiently large $\beta$). 

Figure~\ref{fig:barotropic_cooling} and \ref{fig:non-barotropic_cooling} also 
help to estimate a limit of adiabatic and isothermal approximation. When $\beta>10^2$, 
or up to $10^3$ for lower sound speed, $\eta$ remains to be a constant. This indicates that 
the adiabatic approximation becomes applicable at a limit cooling timescale 
$\beta=10^2\sim 10^3$. Similarly, we find $\eta$ remains unchanged and the isothermal 
limit is approached when $\beta<10^{-1}$. These two limits are both an order of magnitude lager than those proposed by
\cite{2020ApJ...892...65M}. But such a difference is acceptable because the criterion we used 
is different. We mainly focus on the instability grow rate.  However, \cite{2020ApJ...892...65M} determined the limit 
by the angular momentum flux (AMF) of the planet excited waves. 

\subsection{Unstable Growth Rates for Thermal Diffusion Discs}
\begin{figure}
	\includegraphics[width=\columnwidth]{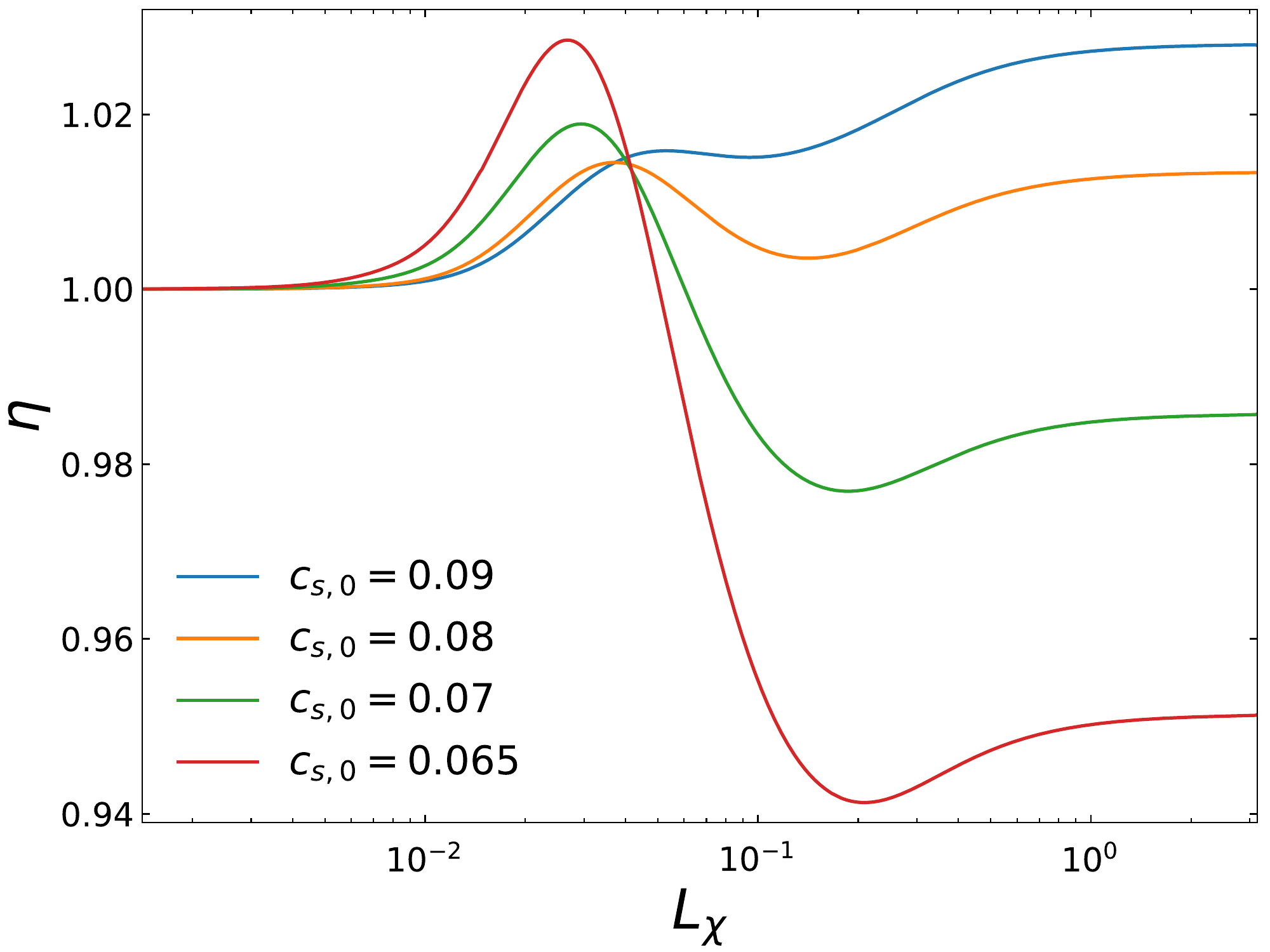}
	\caption{Same as Figure~\ref{fig:barotropic_cooling} but for diffusion discs. 
		Here, the blue, orange, green and red lines represent 
		$c_{s0}= 0.09$, $0.08$, $0.07$, and $0.065$, respectively.}
	\label{fig:diffusion}
\end{figure}

We now turn to discuss the discs with thermal diffusion. 
For a given  diffusivity $\chi$, we solve equations~(\ref{equation_Master_P}) and (\ref{equation_Master_Sigma}) to obtain the growth rate together with the wave functions. The variation of 
the growth rate against diffusivity is shown in Figure~\ref{fig:diffusion}. 
Again, we define a normalized growth rate as  
\begin{equation}
	\eta = \frac{\omega_i}{\omega_i|_{L_\chi=10^{-3}}}, 
	\label{normalization_chi}
\end{equation}
which is the growth rate normalized with the adiabatic limit, and introduce 
\begin{equation}
	L_\chi = \left(\frac{\chi}{\Omega(r_0)}\right)^{\frac12},
\end{equation}
as a length scale of the diffusivity. The lower $L_\chi$ portion of 
Figure~\ref{fig:diffusion} represents adiabatic limit while 
the higher $L_\chi$ portion represents the isothermal limit. 
Our numerical results show that the overall tendencies of $\eta$ variation with different 
diffusivity is similar to those for non-barotropic discs with the simple cooling law. A major distinction is that the growth rate approaches a maximum value between the minimum and adiabatic limit. 
Specifically, the RWI growth rate is enhanced by low diffusivity compared to adiabatic limit.
According to our numerical calculation, the disc approaches a adiabatic limit when $L_\chi$ is lower than  $\sim 10^{-3}$ and an isothermal limit when $L_\chi$ is greater than $\sim 1$.

\subsection{Angular Momentum Flux}
\begin{figure}
	\includegraphics[width=\columnwidth]{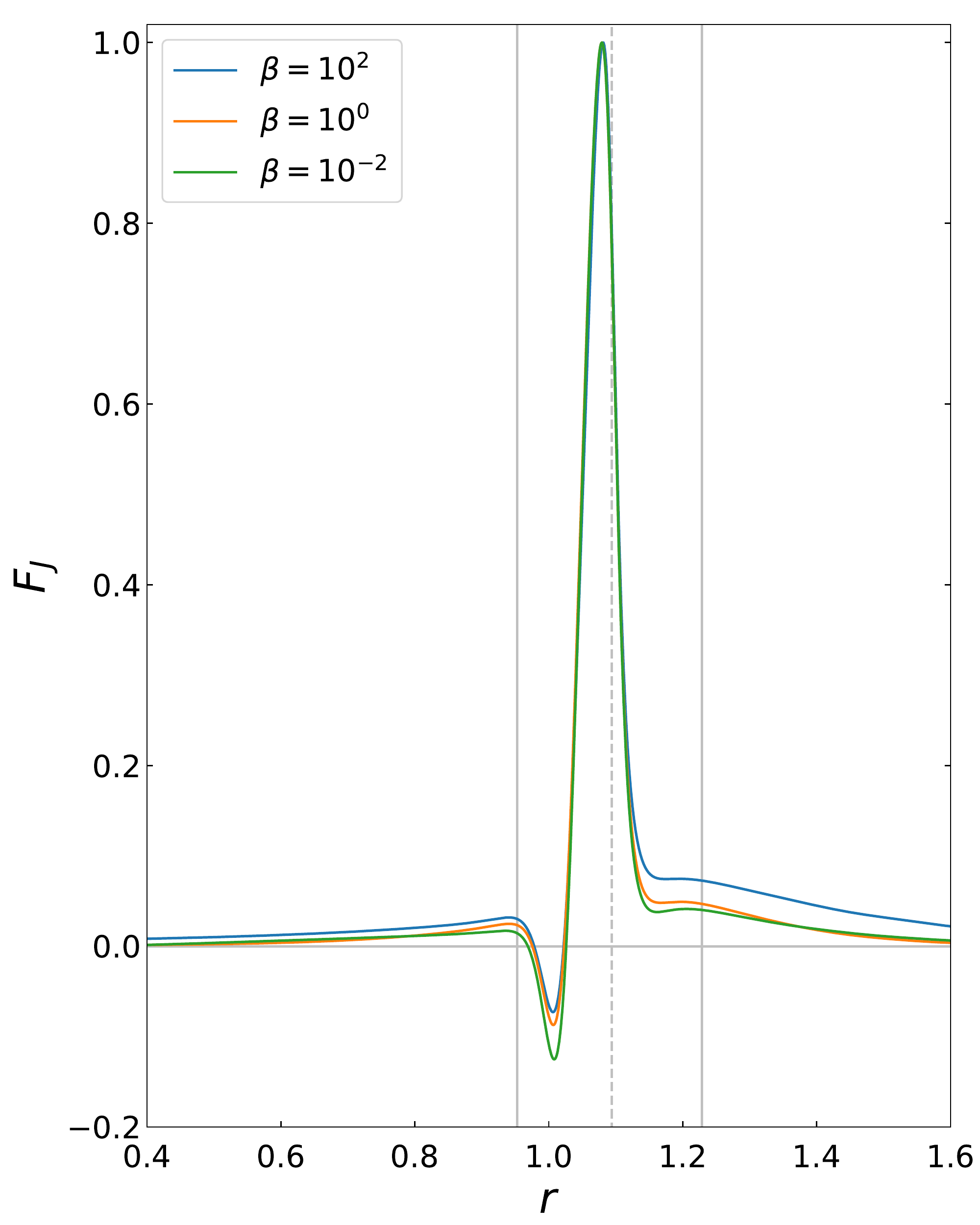}
	\caption{Angular momentum flux for $m=5$, outer edge mode in arbitrary units.
			 different colors represent different cooling timescales. 
			 Orange, green and blue represent $\beta=10^2$,$10^0$ and $10^{-2}$, respectively.
			 The horizontal soild line represents $F_J=0$, the vertical dashed line represents 
			 corotation radius, the left and right vertical lines represent inner and outer 
			 Lindblad resonances radiuses, respectively.}
	\label{fig:AMF}
\end{figure}

It is important to understand the mechanism of angular momentum transfer
by RWI. We can gain physical insight about this by examining the wave angular momentum
flux (AMF),
\begin{equation}
	F_J(r) = r^2\int^{2\pi}_0 d\phi\widetilde{\Sigma} 
	\,\,\widetilde{v}_\phi \widetilde{v}_r.
	\label{equation_F_J}
\end{equation}
In the linear analysis, the right hand side of 
equation~(\ref{equation_F_J}) can be written as 
\citep{1979ApJ...233..857G,2008gady.book.....B,2002ApJ...565.1257T}
\begin{equation}
	F_J(r) = \pi r^2\Sigma\left[
		{\rm Re}(\delta v_r){\rm Re}(\delta v_\phi)
		+{\rm Im}(\delta v_r){\rm Im}(\delta v_\phi)
	\right].
	\label{equation_F_Jm}
\end{equation}
This quantity indicates the total transfer of angular momentum
at radius $r$. The positive values correspond to outward transport and 
negative values correspond to inward transport.
We show AMF associated with the outer edge mode according to equation~(\ref{equation_F_Jm}) for $m=5$  in Figure~\ref{fig:AMF}. 
The $F_J$ is normalized by its own maximum value, which occurs at around the corotation radius. 
The characteristics of these profiles are similar to the result of \cite{2012MNRAS.422.2399M}.
The $F_J$ varies dramatically between the inner Lindblad resonances (ILR) 
and the outer Lindblad resonances (OLR) due to the vortices. 
And it's value at OLR is generally larger than that at ILR, 
which indicate that a net outward angular momentum transfer is carried by the vortices.
Inside the ILR (or outside the OLR), the $F_J$ shows the feature of density wave and remains positive, which indicate outward transport of angular momentum.

The AMF of density waves excited by a planet in cooling discs have been 
discussed by \cite{2019ApJ...878L...9M,2020ApJ...892...65M}.
They found that AMF is conserved, $\frac{\partial F_J}{\partial r}=0$,  
only in adiabatic discs \citep{1979ApJ...233..857G}, and $F_J$ is proportional to $c_s^2$, i.e., $\frac{\partial}{\partial r}\left(\frac{F_J}{c_s^2}\right)=0$, 
in isothermal discs. In addition, the profiles of AMF with different values of 
$\beta$ deviate from those which is conserved, either. But this kind of anomaly 
is not found in AMF driven by RWI in our calculation. 
There is no dramatic changes in the AMF profiles between different cooling time scales in Figure~\ref{fig:AMF}.
The nonlinear hydrodynamic simulation would be beneficial to further understand the angular momentum transport  of RWI in cooling disc.

\section{Conclusions}
\label{sec:conclution}
We perform a linear analysis of RWI in cooling discs. Our calculations
show that vortices are excited by RWI near the  co-rotation resonance. 
These vortices can be considered as potential traps for dust particles and are of crucial importance for the planetesimal formation. 
We concentrate on the RWI growth rate variations in the discs with cooling. The cooling effects are investigated in two different ways,  the simple cooling law as well as the thermal diffusion. 

For the disc with the simple cooling law, we find the RWI growth rate is higher for the longer cooling timescale, while lower for the shorter cooling timescale. The growth rate decreases with the cooling timescales monotonically. However, for non-barotropic discs, the dependence of the growth rates on the cooling timescale in no longer monotonic. There exists a growth rate minimum around $\beta \sim 1$.
\cite{2021ApJ...922...13F} recently investigated the vortices decay with different cooling parameter $\beta$ in the nonlinear regime, and found that a cooling parameter with $\beta\sim 1 - 10$ would lead to the fastest decay. Thus our linear analysis is well consistent with \cite{2021ApJ...922...13F}.
Besides, \cite{2015MNRAS.450.1503L} performed numerical simulation with several 
specific $\beta$ and showed roughly higher growth rate with lower cooling timescale 
in linear region, which is different from our calculations in barotropic discs.
Nevertheless, we obtain similar 
results to \cite{2015MNRAS.450.1503L} for non-barotropic discs with the higher sound speed. Our calculation shows that the cooling effects on RWI is actually more complicated than we expected. Our linear analysis naturally provides a unified framework for various RWI behaviors observed in recent numerical simulations of cooling discs. 

We study the angular momentum flux of RWI in discs with the simple cooling law. The adiabatic and isothermal limits estimated by RWI growth rate are roughly $\beta \sim 10^2 - 10^3$ and $\beta \sim 10^{-1}$, respectively. Such estimations are quite similar to those obtained by criterion according to AMF of density waves  \citep{2020ApJ...892...65M}. The Rossby wave leads to strong angular momentum exchange near co-rotation resonances. 
However, for the density wave propagating in inner discs, we do not find the divergence of AMF indicated by \cite{2019ApJ...878L...9M,2020ApJ...892...65M}. The physical mechanism for this difference is worth further investigations.

When the sound speed of disc is high,  the RWI growth rate for a larger value of $L_{\chi}$ (isothermal limit), is greater than the growth rate for a smaller value of $L_{\chi}$  (adiabatic limit). When the 
disc sound speed is low, the growth rate of isothermal RWI mode is suppressed when compared with the adiabatic RWI mode.
The variations growth rate with $L_{\chi}$ are non-monotonic.
These characteristics are similar to those for non-barotropic discs with simple cooling law. We stress a new feature that is brought about by the thermal diffusion is that the RWI can be even enhanced when the diffusion parameter $L_{\chi}$ lies around $2 - 3 \times 10^{-2}$. 
According to former discussion, if the minimum of growth rate reveals the fast decay of the vortices in nonlinear simulation when $\beta=10^0\sim 10^1$  (\cite{2021ApJ...922...13F,2015MNRAS.450.1503L}), vortices could be boosted or long-term survive with a specific thermal diffusivity relating to this maximum of growth rate.
Apart from this, we estimate a adiabatic and isothermal approximation limit in thermal diffusion discs at around $L_\chi = 10^{-3}$ and $L_\chi = 10^0$, respectively.
But also note that the thermal diffusivity in our study is a constant. It is still unclear that how RWI will vary if the thermal diffusivity is treated as a function of radial, and more precise calculations are still worthwhile.

\section*{Acknowledgements}
We thank the anonymous referee for helpful suggestions that greatly improve this paper. This work has been supported by the National Key R\&D Program of China (No. 2020YFC2201200), the science research grants from the China Manned Space Project (No. CMS-CSST-2021-B09 and CMS-CSST-2021-A10), and opening fund of State Key Laboratory of Lunar and Planetary Sciences (Macau University of Science and Technology) (Macau FDCT Grant No. SKL-LPS(MUST)-2021-2023). C.Y. has been supported by the National Natural Science Foundation of China (Grant Nos. 11373064, 11521303, 11733010, and 11873103).

\section*{Data Availability}
The data underlying this article will be shared on reasonable request to the corresponding author.



\bibliographystyle{mnras}
\bibliography{cooling} 








\bsp	
\label{lastpage}
\end{document}